\documentclass[journal,10pt, final,a4paper]{IEEEtran}

\usepackage{cite}

\ifCLASSINFOpdf
   \usepackage[pdftex]{graphicx}
\else
\fi

\usepackage{amsmath}

\usepackage{wrapfig}

\usepackage[caption=false]{subfig}
\usepackage{graphicx}

\usepackage[latin1]{inputenc}
\usepackage{amssymb}
\usepackage{multirow}
\usepackage{rotating}
\usepackage{picins}
\usepackage{floatflt}
\usepackage{url}
\usepackage{hyperref}
\usepackage{algorithm}
\usepackage{algpseudocode}
\usepackage{pifont}
\newcommand{\cmark}{\ding{51}}%
\newcommand{\xmark}{\ding{55}}%
\newcommand{\comment}[1]{}

\usepackage{array}
\usepackage{arydshln}
\usepackage{booktabs}
\newcommand{\ra}[1]{\renewcommand{\arraystretch}{#1}}
\usepackage{color}

\newcommand{\rebuttal}{\textcolor{black}}

\begin{document}
%

 \title{ Recalibrating 3D ConvNets with Project \& Excite}

\author{Anne-Marie~Rickmann,
Abhijit~Guha~Roy,
Ignacio~Sarasua,
and~Christian~Wachinger 

\thanks{ 
A. Rickmann, A. Guha Roy, I. Sarasua and C. Wachinger are with the Lab for Artificial Intelligence in Medical Imaging (AI-Med), Department of Child and Adolescent Psychiatry, University Hospital, Ludwig-Maximilians-University M\"{u}nchen, Germany. (E-mail: arickman@med.lmu.de (A. Rickmann))
}
}

\markboth{Preprint submitted to Transactions on Medical Imaging}
{Rickmann \MakeLowercase{\textit{et al.}}: Recalibrating 3D ConvNets}

\maketitle

\begin{abstract}
Fully Convolutional Neural Networks (F-CNNs) achieve state-of-the-art performance for segmentation tasks in computer vision and medical imaging. 
Recently, computational blocks termed \emph{squeeze and excitation} (SE) have been introduced to recalibrate F-CNN feature maps both channel- and spatial-wise, boosting segmentation performance while only minimally increasing the model complexity. 
So far, the development of SE blocks has focused on 2D architectures. 
For volumetric medical images, however, 3D F-CNNs are a natural choice. 
In this article, we extend existing 2D recalibration methods to 3D and propose a generic \emph{compress-process-recalibrate} pipeline for easy comparison of such blocks.
We further introduce \emph{Project \& Excite} (PE) modules, customized for 3D networks.
In contrast to existing modules, Project \& Excite does not perform global average pooling but compresses feature maps along different spatial dimensions of the tensor separately to retain more spatial information that is subsequently used in the excitation step.
We evaluate the modules on two challenging tasks, whole-brain segmentation of MRI scans and whole-body segmentation of CT scans.  
We demonstrate that PE modules can be easily integrated into 3D F-CNNs, boosting performance up to 0.3 in Dice Score and outperforming 3D extensions of other recalibration blocks, while only marginally increasing the model complexity.
Our code is publicly available on 
\href{https://github.com/ai-med/squeeze_and_excitation}{https://github.com/ai-med/squeeze\_and\_excitation}.
\end{abstract}


\IEEEpeerreviewmaketitle


\section{Introduction}
\label{sec:intro}
Fully convolutional neural networks (F-CNNs) have been widely adopted for semantic image segmentation in computer vision~\cite{long2015fully,chen2017deeplab,badrinarayanan2017segnet} and medical image analysis~\cite{ronneberger2015u,roy2019quicknat}.
Most of the architectural innovations focus on 2D CNNs, as computer vision tasks mainly deal with 2D natural images. 
These innovations are often not directly applicable for processing volumetric medical scans like CT, MRI, and PET. 
2D F-CNNs are typically used to segment 3D medical scans slice-wise. In such an approach, the contextual information from adjacent slices remains unexplored, which might lead to imperfect segmentations~\cite{milletari2017hough}. 
Hence, the natural choice for segmenting 3D scans are 3D F-CNN architectures.
3D F-CNNs for medical image segmentation have become more popular in recent years and have shown promising results~\cite{cciccek20163d,milletari2016v,wachinger2018deepnat,Chen2018,dolz2018hyperdense}.
However, there exist practical challenges in using 3D F-CNNs. The number of learnable weight parameters is much higher than for their 2D counterparts, which makes these models prone to overfitting when training data is limited. Furthermore, they require a large amount of GPU RAM for training.
The first challenge is particularly pronounced in medical image segmentation, where most benchmark datasets generally consist of only 15-20 labeled scans~\cite{landman2012miccai, visceral}.
Highly complex 3D F-CNNs are susceptible to overfitting when trained with such limited data.
This problem is commonly mitigated by carefully engineering 3D F-CNNs for a particular task by minimizing the model complexity conditioned on the amount of available training data. This can be done by either reducing the number of convolutional layers or by decreasing the number of channels per convolutional layer.
Although reducing model complexity can aid training models with limited data, the exploratory capacity of the 3D F-CNN gets limited.
The second challenge of limited memory is commonly addressed by partitioning the full volume into subvolumes and training on them instead. 
The disadvantage is, however, that the context of the model is reduced, similar to 2D F-CNNs, and strategies are required for stitching the full volume back together~\cite{huo2018spatially}. 
Concluding, for 3D F-CNNs it is necessary to ensure that the learnable parameters within the network are effectively utilized, creating a need for methods that aid the network in learning useful features without further increasing model complexity.

Recently, Hu et al.~\cite{Hu_2018_CVPR} introduced a computational module termed `Squeeze and Excite' (SE) to recalibrate CNN feature maps, which boosts the performance while only marginally increasing model complexity. 
SE blocks model the interdependencies between the channels of feature maps and learn to provide attention on specific channels depending on the task. 
Channel interdependencies are learned by first squeezing the spatial information channel-wise through average pooling and secondly passing the vector of channel descriptors through a fully connected subnetwork to learn channel-specific weights. 
The input feature map is then scaled by the weights and therefore channels can be selectively emphasized or suppressed.
Hu et al. demonstrated the ease of including SE modules into state-of-the-art 2D CNN architectures, providing a boost in performance on classification tasks with a fractional increase in learnable parameters.

\begin{figure}[t]
    \centering
     \includegraphics[width= 0.4 \textwidth]{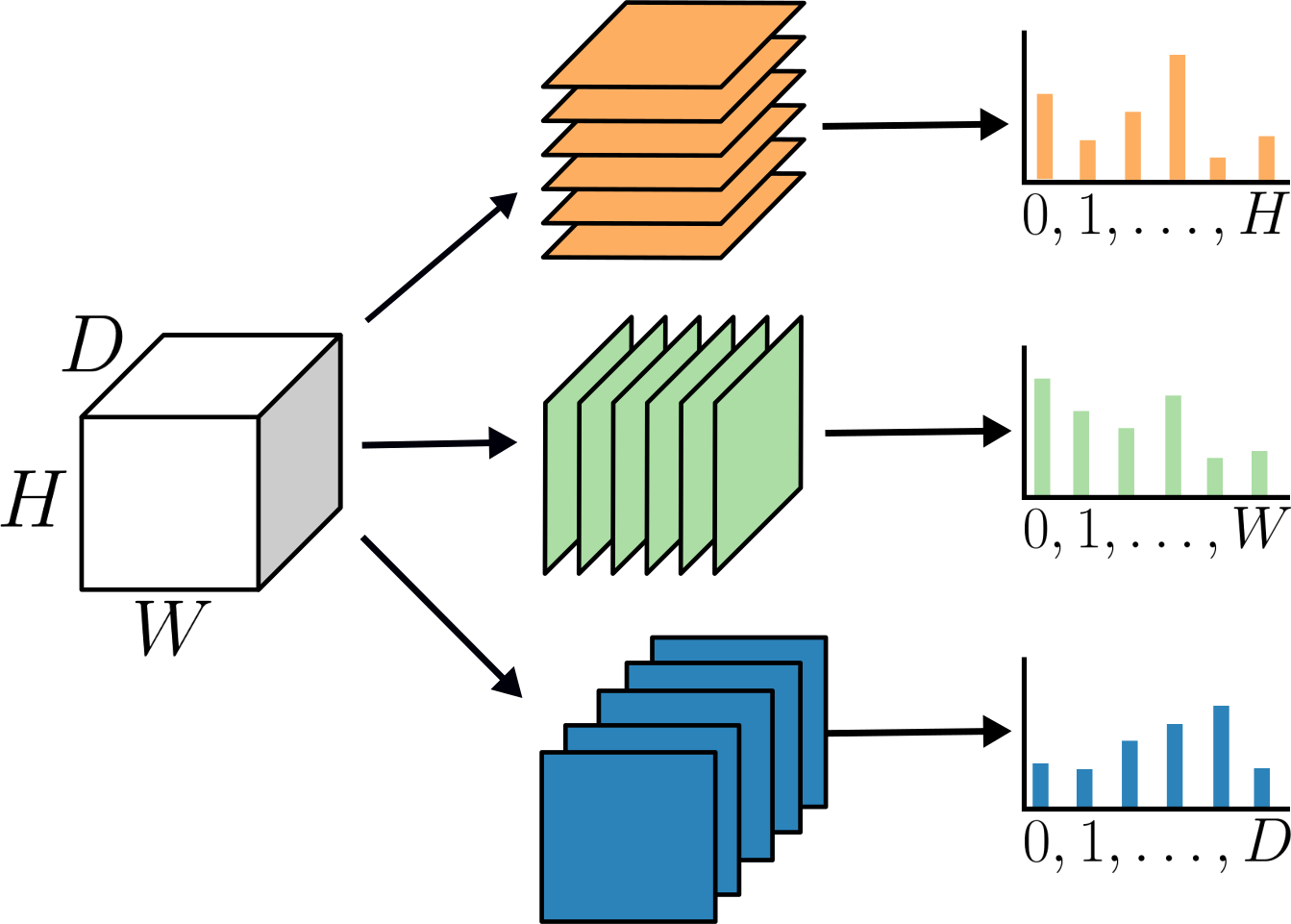}
    \caption{Slicing of a 3D tensor among the three axes and 1D projections of the slices, e.g., calculated by average pooling of each slice as used in Project \& Excite blocks.}
    \label{fig:motivation}
\end{figure}

In this article, we study the recalibration of feature maps within 3D F-CNNs with different recalibration blocks. In particular, we introduce the `Project \& Excite' (PE) module, a new computational block custom-made for 3D inputs.
We hypothesize, that removing all spatial information of a high-dimensional feature map by global pooling, as in SE, leads to a loss of relevant information, particularly for segmentation, where we need to exactly localize anatomical structures. 
In contrast, we aim at preserving the spatial information while still providing a global receptive field to the network at every stage.
We draw our inspiration from traditional tensor slicing techniques~\cite{rabanser2017introduction}, by averaging along the three principal axes of the tensor as illustrated in Fig.~\ref{fig:motivation}.
By this, we obtain three projection-vectors indicating the relevance of the slices along the three axes. A spatial location is important if all the corresponding slices associated with it provide higher estimates.
So, instead of learning dependencies of scalar values across the channels, as in SE, we learn the dependencies of these projection-vectors across the channels for excitation.

This article extends our earlier work~\cite{rickmann2019project} by providing a generic framework for recalibration methods, comparing our method with 3D extensions of other recalibration blocks, integrating PE blocks into different architectures and validation of the module across different datasets. 
The main contents of this article are: 
\begin{enumerate}
    \item We propose a new computational block termed `Project \& Excite' for recalibration of 3D F-CNNs.
    \item We show that our proposed PE block can be easily integrated into 3D F-CNNs, by including them into two different F-CNN architectures. 
    \item We demonstrate that PE blocks minimally increase the model complexity compared to using more convolutional layers, while providing higher segmentation accuracy, especially for small target classes. 
    \item We introduce the compress-process-recalibrate pipeline for easy comparison of recalibration blocks.
    \item We provide 3D extensions to existing recalibration techniques and compare them with our PE blocks, supporting our hypothesis that preserving more spatial information is crucial in 3D settings.
\end{enumerate}{}

\subsection{Related work}

Two of the earlier 3D architectures, 3D U-net~\cite{cciccek20163d} and V-net~\cite{milletari2016v}, are based on the encoder-decoder structure of 2D U-net~\cite{ronneberger2015u}. In both networks, the number of channels per convolutional layer is much smaller than in a typical 2D network. 
A common strategy to alleviate the problem of high memory demand of 3D networks is to train 3D F-CNNs on subvolumes or image segments~\cite{kamnitsas2017efficient,wachinger2018deepnat,dolz2018hyperdense,Chen2018}.
When training on subvolumes, it is important to choose the sampling strategy according to the given task. 
Some lighter networks can pass a whole volume during inference since there is no need to store activations for backpropagation, which leads to a full volume segmentation map. Other networks process the volume in segments also during inference. If the segments are overlapping, they have to be stitched to obtain a full volume segmentation and a label fusion strategy for overlapping sections is needed.
Huo et al.~\cite{huo2018spatially} propose to divide a brain volume into overlapping subspaces, register the subvolumes to a common atlas, and train a separate 3D F-CNN for each subspace. 
In contrast to the common strategy of training on subvolumes, we aim at training 3D-FCNNs on full volumes, with no need for additional pre- or postprocessing and cumbersome stitching methods. 

Various authors have extended Squeeze and Excitation modules and applied them to different classification and segmentation tasks in computer vision and medical image analysis.
Roy et al.~\cite{roy2019recalibrating} extended the idea of SE to medical image segmentation and introduced a recalibration block called spatial squeeze and excite (sSE).
The idea is that for segmentation tasks the fine-grained spatial information is highly important and therefore needs to be preserved. The sSE block squeezes channel information and performs the recalibration spatially. 
The authors demonstrated that sSE outperforms the original channel SE module (cSE)~\cite{Hu_2018_CVPR} for medical segmentation tasks, while a combination of both modules (scSE) reaches the highest performance. They demonstrated that such light-weight blocks can be a better architectural choice than extra convolutional layers. 

The convolutional block attention module~(CBAM)~\cite{woo2018cbam} combines channel and spatial attention modules in a sequential manner. Both modules are similar to squeeze and excitation blocks and include a combination of max and average pooling for squeezing the channel and spatial information. The authors showed that including max-pooling increased the performance compared to using average or max-pooling separately. 
Although sSE, scSE, and CBAM have shown promising results on 2D segmentation tasks, the efficiency of 3D extensions of these modules has not yet been evaluated.

Pereira et al.~\cite{pereira2019adaptive} introduced a jointly learned channel and spatial recalibration module, termed SegSE, for medical segmentation tasks based on dilated convolutions instead of average pooling. Although they showed their module performs better than recalibration using pooling methods, dilated convolutions come with a higher demand for GPU memory. 

Although cSE blocks were customarily designed for 2D architectures, they have recently been extended for 3D F-CNNS to aid volumetric segmentation~\cite{zhu2019anatomynet}. Zhu et al. directly extended the cSE module to 3D to perform channel recalibration, applied to medical image segmentation. They showed an improved performance over baseline models without recalibration blocks, but they do not perform spatial recalibration.

\begin{table*}[h]
    \centering
     \caption{Comparison of squeeze and excite variations and our proposed Project \& Excite module with respect to the compress ($\mathbf{C}(\cdot)$), process  ($\mathbf{P}(\cdot)$) and recalibrate ($\mathbf{R}(\cdot,\cdot)$) operations. The second column shows for which type of CNN (2D or 3D) the module has been previously used.}
     \ra{1.2}
    \begin{tabular}{@{}l c c c c c c c @{} }
        \toprule
          & & \multicolumn{2}{c}{$\mathbf{C}(\cdot)$} & \multicolumn{2}{c}{$\mathbf{P}(\cdot)$} &\multicolumn{2}{c}{$\mathbf{R}(\cdot,\cdot)$}\\
        \cmidrule(lr{.75em}){3-4}\cmidrule(lr{.75em}){5-6} \cmidrule(lr{.75em}){7-8}
        Module & Used in & Linear & Parametric & FC & Conv & Gating function & Recalibration\\
        \midrule
         cSE~\cite{Hu_2018_CVPR,zhu2019anatomynet} & 2D \& 3D CNNs & \cmark & \xmark & \cmark & \xmark & sigmoid & channel-wise multiplication\\
         sSE~\cite{roy2019recalibrating} & 2D CNNs & \cmark &\cmark & \xmark & \cmark & sigmoid & element-wise multiplication\\
         CBAM channel~\cite{woo2018cbam} & 2D CNNs & \xmark & \xmark & \cmark & \xmark & sigmoid & channel-wise multiplication\\
         CBAM spatial~\cite{woo2018cbam} & 2D CNNs& \xmark& \xmark & \xmark & \cmark & sigmoid & element-wise multiplication\\
         Project \& Excite & 3D CNNs & \cmark & \xmark & \xmark & \cmark & sigmoid & element-wise multiplication\\
         \bottomrule
    \end{tabular}
    \label{tab:cpr}
\end{table*}{}

\section{Methods}

\begin{figure}[t]
  \centering
  \includegraphics[width=\linewidth]{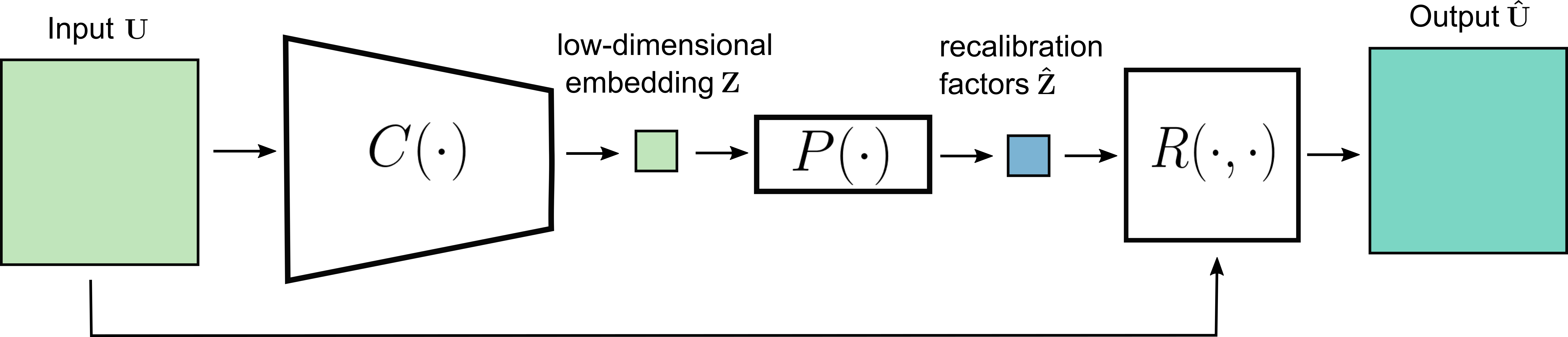}
  \caption{Illustration of the CPR framework. An input feature map $\mathbf{U}$ is passed through the \textit{Compressor} $\mathbf{C}(\cdot)$ and compressed to a lower-dimensional embedding $\mathbf{Z}$. Recalibration factors $\hat{\mathbf{Z}}$ are learned by the \textit{Processor} $\mathbf{P}(\cdot)$ and finally the input feature map gets scaled by $\hat{\mathbf{Z}}$ in the \textit{Recalibration} function $\mathbf{R}(\cdot,\cdot)$, yielding the output $\mathbf{\hat{U}}$  }
\label{fig:cpr_visual}
\end{figure}

Previously introduced recalibration blocks and our proposed Project \& Excite module follow a similar procedure for recalibration. 
To facilitate the comparison of these different methods, we present a generic framework that we call \textit{compress-process-recalibrate} (CPR). 
All recalibration blocks take a high dimensional feature map, usually the output of a previous convolutional layer within the network, as input.
First, the function $\mathbf{C}(\cdot)$ compresses the high dimensional input feature map $\mathbf{U}$ to a lower-dimensional embedding $\mathbf{Z}$. Then, the processor $\mathbf{P}(\cdot)$ learns a mapping from the low dimensional embedding $\mathbf{Z}$ to recalibration factors $\mathbf{\hat{Z}}$.
The final recalibration step $\mathbf{R}(\cdot,\cdot)$ first rescales $\mathbf{\hat{Z}}$ by a gating function and finally scales the input feature map with $\hat{\mathbf{Z}}$, yielding the output feature map $\hat{\mathbf{U}}$, where channels or spatial locations get emphasized or suppressed.
We provide a schematic illustration of the CPR framework in Fig.~\ref{fig:cpr_visual}.
There are several ways to compress the feature map using linear or non-linear pooling operations, which are inherently non-parametric or using parametric operations like convolutions.
The processor is usually parametric by using either fully connected or convolutional subnetworks. Tab.~\ref{tab:cpr} characterizes various recalibration blocks in the CPR framework. 
In the following, we will detail existing recalibration blocks, extend them to 3D and finally introduce the 'Project \& Excite' block.

\subsection{3D channel Squeeze \& Excite}
The only existing 3D SE block~\cite{zhu2019anatomynet} is a direct extension of the original 2D SE block~\cite{Hu_2018_CVPR}, which we refer to as 3D channel SE (cSE) module. 

The compressor $\mathbf{C}: \mathbb{R}^{H\times W\times D\times C} \rightarrow \mathbb{R}^{C}$ performs a global average pooling operation that squeezes the spatial content of the input $\mathbf{U}$  into a scalar value per channel $\mathbf{z} \in \mathbb{R}^{C}$, hence the name 'squeeze' in the original version. For simplicity, we describe a single channel of the input $\mathbf{U}$ as $\mathbf{u}_c$.
The processor operation $\mathbf{P}(\cdot)$ takes in $\mathbf{z}$ and adaptively learns the inter-channel dependencies by using two fully connected layers. 
In the recalibration function $\mathbf{R}(\cdot,\cdot)$, the activations $\mathbf{\hat{z}}$ are passed through a sigmoid gating function to ensure that multiple channels can be emphasized or suppressed. Finally, the input feature map gets scaled with the learned recalibration weights channel-wise.
The operations are defined as:
\begin{align}
    \mathbf{C}&: \quad \mathbf{z}  = \text{AvgPool}(\mathbf{U}), \\
    \mathbf{P}&: \quad \mathbf{\hat{z}}  = \mathbf{W}_2\delta(\mathbf{W}_1 \mathbf{z}),\\
    \mathbf{R}&: \quad \mathbf{\hat{u}}_c = \sigma(\hat{z}_c) \mathbf{u}_c,
\end{align}
\noindent
with AvgPool describing the channel-wise average pooling operation, $\delta$ denoting the ReLU nonlinearity, $\sigma$ the sigmoid layer, $\mathbf{W}_1 \in \mathbb{R}^{\frac{C}{r}\times C}$ and $\mathbf{W}_2 \in \mathbb{R}^{C \times \frac{C}{r}}$ the weights of the fully connected layers. The hyperparameter $r$ is the channel reduction factor similar to~\cite{Hu_2018_CVPR}, which allows us to adjust the computational and memory cost of the cSE block.

\subsection{3D spatial Squeeze and Excite}
The spatial SE block (sSE)~\cite{roy2019recalibrating} was designed specifically for segmentation tasks.
We provide an extension to 3D by replacing all functions by their 3D counterparts. 
Contrary to the other modules, sSE includes the \textit{process} step in the \textit{compress} transformation.
$\mathbf{C,P}$ and $\mathbf{R}$ are defined as:
\begin{align}
    \mathbf{C},\mathbf{P}&: \quad \mathbf{Z} =  \mathbf{S} \star \mathbf{U},\\ 
    \mathbf{R}&: \quad \mathbf{\hat{u}_c} =  \sigma(\mathbf{Z}) \cdot \mathbf{u}_c ,
\end{align}
where $\mathbf{S} \in \mathbb{R}^{1\times1\times1\times C \times 1}$ are the weights of the convolution kernel.
The compressor operation compresses channel information by using a $1\times1\times1$ kernel to reduce the channel dimension to $1$. 
The resulting recalibration map is rescaled by a sigmoid layer and multiplied with each channel of the input feature map element-wise in the recalibration operation. 

\subsection{3D spatial and channel Squeeze \& Excite}
The combination of cSE and sSE blocks has been proposed in~\cite{roy2019recalibrating}, referred to as spatial and channel SE (scSE).
In this block, the input feature map $\mathbf{U}$ is passed through a cSE and sSE block separately. The two output feature maps $\mathbf{\hat{U}}_{cSE}$ and $\mathbf{\hat{U}}_{sSE}$ are then combined by an element-wise max operation to obtain the final output $\mathbf{\hat{U}}_{scSE}$. We obtain a 3D extension by combining the previously described 3D cSE and 3D sSE blocks.
\begin{figure*}[h]
  \centering
  \includegraphics[width=0.75\linewidth]{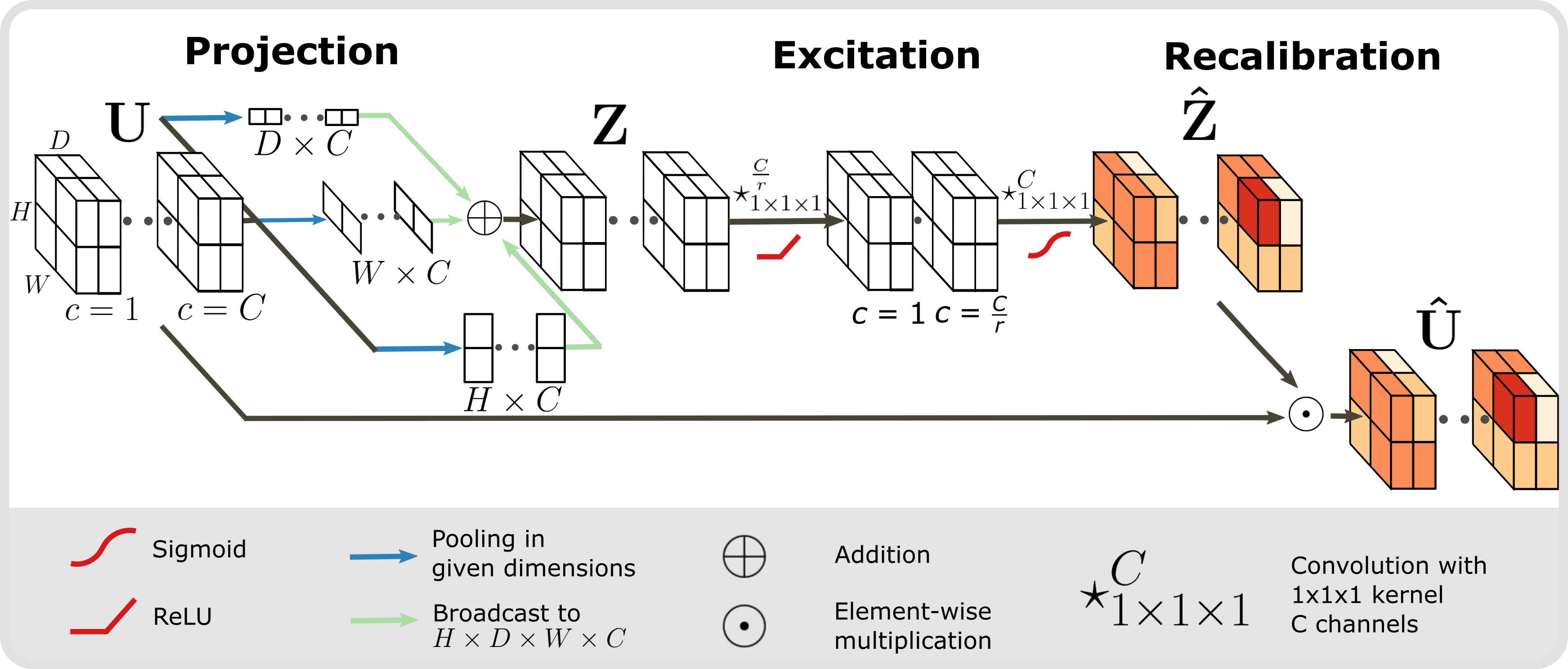}
  \caption{Illustration of the proposed 'Project\& Excite' block. On the left, the projection operation, that takes a 4D tensor U (with width \textit{W}, height \textit{H}, depth \textit{D} and channels \textit{C}) as input and calculates the 3 projection vectors by pooling operations. On the right, the excitation operation with 2 convolutional layers and the recalibration of the feature map. The reduction rate hyper-parameter is indicated by $r$.}
\label{fig:pe_block}
\end{figure*}
\subsection{3D Convolutional Block Attention Module}
The convolutional block attention module (CBAM)~\cite{woo2018cbam} was designed for 2D classification and object detection tasks. To the best of our knowledge, this block has not yet been used for 3D segmentation tasks. Due to its similarity to squeeze and excite blocks we chose to compare to a 3D version of CBAM. 
CBAM is divided into a channel and spatial attention block, which are combined sequentially. 
The channel attention block is similar to 3D cSE. The compressor performs global max and average pooling, with the result being passed through a shared fully connected subnetwork.
The features are merged by adding them element-wise, before passing them through a sigmoid gating function. Finally, the input feature map is scaled by element-wise multiplication with the learned weights. The $\mathbf{C}, \mathbf{P},$ and $\mathbf{R}$ functions are defined as:
\begin{align}
    \mathbf{C}_{avg}&: \quad \mathbf{z}_{avg} = \text{AvgPool}( \mathbf{U}), \\
    \mathbf{C}_{max} &: \quad \mathbf{z}_{max} = \text{MaxPool}(\mathbf{U}),\\
    \mathbf{P} &:\quad \mathbf{\hat{z}}= \mathbf{W}_2\delta(\mathbf{W}_1 \mathbf{z}_{avg}) + \mathbf{W}_2\delta(\mathbf{W}_1 \mathbf{z}_{max}),\\
    \mathbf{R}&: \quad \mathbf{\hat{u}}_c = \sigma(\hat{z}_c) \mathbf{u}_c,
\end{align}{}
with AvgPool and MaxPool denoting the channel-wise pooling operations, and $\mathbf{W}_1 \in \mathbb{R}^{\frac{C}{r}\times C}$ and $\mathbf{W}_2 \in \mathbb{R}^{C \times \frac{C}{r}}$ denoting the weights of the fully connected layers.

The spatial attention block compresses channel information by performing average pooling and max pooling along the channel dimension and concatenates the resulting descriptors along the channel dimension. The concatenated descriptor is passed through a $1\times1\times1$ convolutional layer followed by a sigmoid layer to generate the spatial attention map. $\mathbf{C}(\cdot)$, $\mathbf{P}(\cdot, \cdot)$ and $\mathbf{R}(\cdot, \cdot)$ are defined as:
\begin{align}
    \mathbf{C}_{avg}&: \quad \mathbf{Z}_{avg} = \text{AvgCPool}(\mathbf{U})\\
    \mathbf{C}_{max}&: \quad \mathbf{Z}_{max} = \text{MaxCPool}(\mathbf{U})\\
    &\quad \quad \mathbf{Z} = [\mathbf{Z}_{avg};\mathbf{Z}_{max}]\\
    \mathbf{P}&: \quad \mathbf{\hat{Z}} = \mathbf{V} \star \mathbf{Z}\\
    \mathbf{R}&: \quad \mathbf{\hat{U}} = \mathbf{\hat{Z}} \cdot \mathbf{u}_c,
\end{align}{}
with AvgCPool($\cdot$) and MaxCPool($\cdot$) denoting the channel-wise average and max pooling operations, $[\cdot ; \cdot]$ the concatenation in the channel dimension, $\star$ the convolution operation and $\mathbf{V} \in \mathbb{R}^{1\times 1\times 1\times 2 \times 1}$ the convolutional weights.
The blocks are combined sequentially by passing the input through the channel attention block first and then passing the result through the spatial attention block.

\subsection{`Project \& Excite' Module}
\label{sec:methods:PE}
The previously described recalibration blocks have been designed for 2D tasks and their direct 3D extensions might therefore not be optimal for 3D segmentation tasks.
The cSE, scSE and CBAM blocks compress spatial information of a volumetric feature map into one scalar value per channel. Especially in the first/last layers of a typical encoder-decoder architecture, these feature maps have a high spatial extent. 
We hypothesize, that a global pooling operation might not properly capture the relevant spatial information of a large-sized 3D input.
Hence, we introduce the `Project \& Excite' module that retains more of the valuable spatial information within our proposed projection operation. The excitation operation then learns inter-dependencies between the projections across the different channels. Thus, it combines spatial and channel information for recalibration. 
Fig.~\ref{fig:pe_block} illustrates the architecture of the `PE' block.
The projection operation $\mathbf{C}(\cdot)$ is separated into three projection operations ($\mathbf{C}_{H}(\cdot)$, $\mathbf{C}_{W}(\cdot)$, $\mathbf{C}_{D}(\cdot)$) along the spatial dimensions with outputs $\mathbf{z}_{h_c} \in \mathbb{R}^{C \times H}$, $\mathbf{z}_{w_c} \in \mathbb{R}^{C \times W}$, and $\mathbf{z}_{d_c} \in \mathbb{R}^{C \times D}$. 
The projection operation can be defined as any pooling operation. Here we describe averaging along the spatial dimensions as an example:
\begin{align}
    \mathbf{C}_{H}&: \quad \mathbf{z}_{h_c}(i) = \frac{1}{W}\frac{1}{D}\sum_{j=1}^W\sum_{k=1}^D \mathbf{u}_c(i,j,k), \\
    \mathbf{C}_{W}&: \quad \mathbf{z}_{w_c}(j) = \frac{1}{H}\frac{1}{D} \sum_{i=1}^{H}\sum_{k=1}^D \mathbf{u}_c(i,j,k),\\
    \mathbf{C}_{D}&: \quad \mathbf{z}_{d_c}(k) = \frac{1}{H}\frac{1}{W} \sum_{i=1}^{H}\sum_{j=1}^W \mathbf{u}_c(i,j,k), 
\end{align}
with $i \in \{1, \ldots, H\},  \quad j \in \{1, \ldots, W\}, \quad k \in \{1, \ldots, D\}$.
The outputs $\mathbf{z}_{h_c}, \mathbf{z}_{w_c}$, and  $\mathbf{z}_{d_c}$ are broadcasted to the shape $H \times W \times D \times C$ and added to obtain $\mathbf{Z}$, which is then fed to the processor  $\mathbf{P}(\cdot)$. The processor is defined by two convolutional layers around a ReLU activation.
The convolutional layers have kernel size $1 \times 1 \times 1$, to aid the modeling of channel dependencies. The first layer reduces the number of channels by $r$, and the second layer brings the channel dimension back to the original size. The process and recalibrate operations are defined as:
\begin{align}
    \mathbf{P}&: \quad \hat{\mathbf{Z}} = \mathbf{V}_2 \star \delta (\mathbf{V}_1 \star \mathbf{Z}),\\
    \mathbf{R}&: \quad \hat{\mathbf{U}} =  \sigma(\hat{\mathbf{Z}}) \odot \mathbf{U},
\end{align}
where $\star$ describes the convolution operation, $\odot$ indicates point-wise multiplication, $\mathbf{V}_1 \in \mathbb{R}^{1 \times 1 \times 1 \times \frac{C}{r}}$ and $\mathbf{V}_2 \in \mathbb{R}^{1 \times 1 \times 1 \times C}$ the convolution weights.
The final output of the PE block $\hat{\mathbf{U}}$ is obtained by an element-wise multiplication of the feature map $\mathbf{U}$ and $\hat{\mathbf{Z}}$. 
\subsection{Integration into F-CNN architectures}
\begin{figure}[b]
    \centering
    \includegraphics[width=\linewidth]{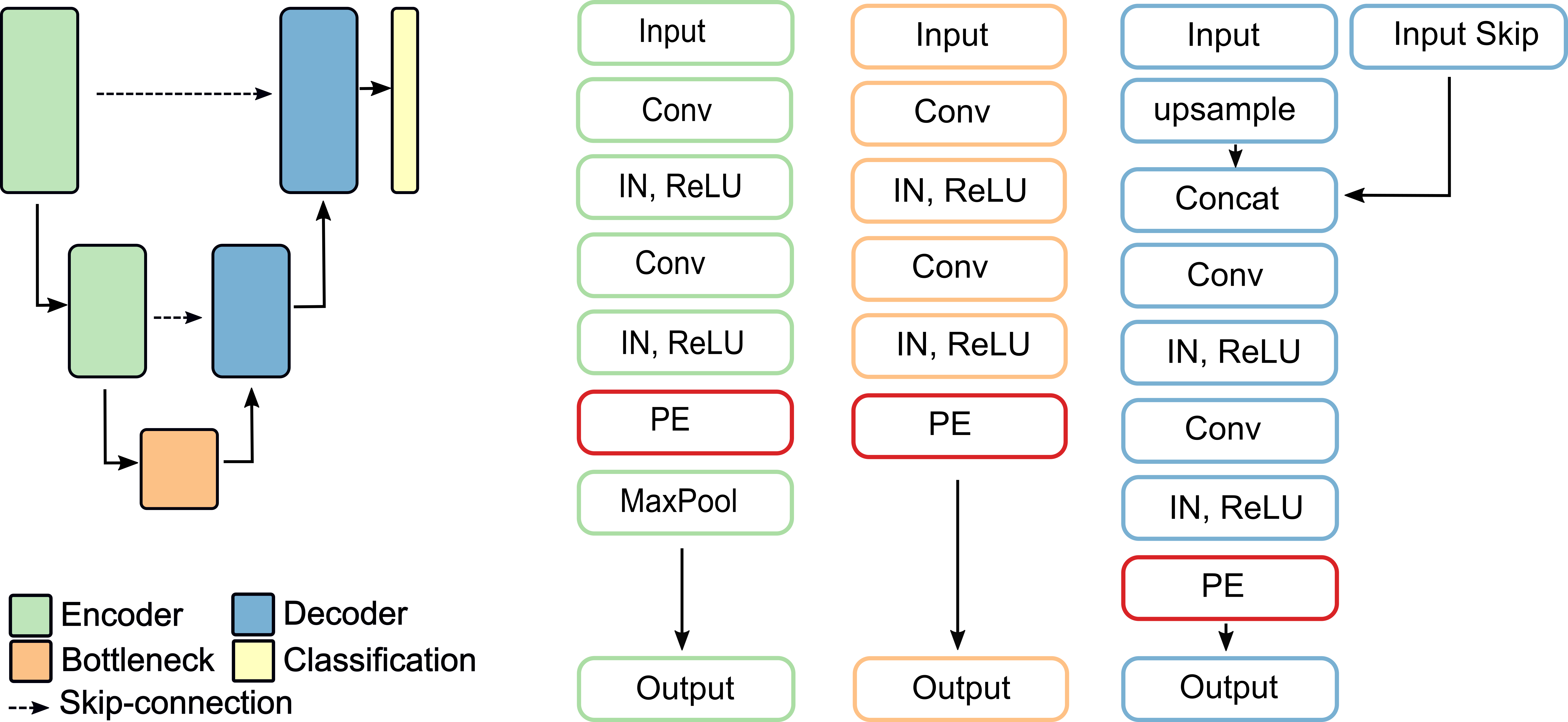}
    \caption{Left: Schematic illustration of 3D U-net with encoder, decoder, bottleneck, and classification blocks. Right: We show exemplary the placement of PE blocks within an encoder, decoder and bottleneck block, respectively. 'IN' indicates instance normalization as used in our experiments. }
    \label{fig:placement_U-net}
\end{figure}

Recalibration blocks can be easily integrated into existing F-CNN architectures. They are typically placed after the non-linearity following a convolutional layer~\cite{Hu_2018_CVPR}. We follow the same strategy with our 3D extensions and PE modules. We illustrate possible placements of PE blocks within a typical encoder-decoder based network in Fig.~\ref{fig:placement_U-net} and validate these choices in Sec.~\ref{section:position}. Hu et al. also successfully integrated cSE blocks within residual networks. We investigate the performance of 3D recalibration blocks within a residual 3D F-CNN in Sec.~\ref{section:exp_voxresnet}.

\section{Experimental Setup}

\subsection{Datasets}
We chose two challenging 3D segmentation tasks: Whole-brain segmentation of MRI scans and Whole-body segmentation of ceCT scans. 
Both tasks involve segmentation of a substantial number of target classes with highly variable shape and size introducing very high class-imbalance. 
The details of the used dataset and splits are provided below.

\subsubsection{Whole-brain segmentation of MRI T1 scans}
In our experiments, we use three different brain MRI datasets. We segment these brain scans into 32 cortical and subcortical structures. We use the MALC dataset for training and ADNI, and CANDI datasets for testing. The manual annotations for all brain datasets were provided by Neuromorphometrics, Inc.
\paragraph{MALC Dataset}
The Multi-Atlas Labelling Challenge (MALC) dataset~\cite{landman2012miccai} is part of the OASIS dataset~\cite{marcus2010open}. It consists of $30$ T1 MRI volumes of the brain, each from a different subject. All scans have an isotropic resolution of $1\text{mm}^3$. We use this dataset for training the model. Due to the limited data, we perform a 5-fold cross-validation, using $24$ scans for training and $6$ scans for testing in each fold. \rebuttal{During training, $2$ scans of the training set were kept as a validation set.}
\paragraph{ADNI-29 Dataset}
The dataset consists of 29 scans from the ADNI dataset~\cite{jack2008alzheimer}, with a balanced distribution of Alzheimer's Disease and control subjects, and scans acquired with 1.5T and 3T scanners. Presence of pathology makes the segmentation task challenging.
\paragraph{CANDI Dataset}
The dataset consists of 13 brain scans of children (age 5-15) with psychiatric disorders and is part of the CANDI dataset~\cite{kennedy2012candishare}. Some scans have severe motion artifacts.

\subsubsection{Whole-body segmentation of ceCT scans}
In this experiment, we use the contrast-enhanced whole-body CT scans from the Visceral dataset~\cite{visceral}.
The dataset consists of $20$ annotated scans with a voxel resolution of $2\text{mm}^3$.   
We segment $14$ organs from the thorax and abdomen.
We perform 5-fold cross-validation, where one scan from the test fold was kept as the validation set.
\rebuttal{We perform 5-fold cross-validation, with $16$ scans for training and $4$ scans for testing in each fold. During training, $2$ scans of the training set are kept as validation set.}

\subsection{Baseline Architectures}
The three most commonly used 3D F-CNN architectures are 3D U-net~\cite{cciccek20163d}, V-net~\cite{milletari2016v} and VoxResNet~\cite{Chen2018}. 3D U-net and V-net both have a similar encoder-decoder skeleton whereas VoxResNet has a different architecture with side supervision. 
In this paper, we chose to evaluate our proposed PE blocks on one encoder-decoder architecture (3D U-net) and one side supervision architecture (VoxResNet). 

\subsubsection{3D U-net}
3D U-net is a typical segmentation network, with an encoding and decoding path, connected with skip connections. The network architecture is schematically illustrated in Fig.~\ref{fig:placement_U-net}.
We reduced the number of parameters to ensure proper trainability on whole 3D scans.
Our design consists of 3 encoder and 3 decoder blocks, with only the first two encoders performing downsampling, and the last two decoders performing upsampling. Each encoder/decoder consists of 2 convolutional layers with kernel size $3\times3\times3$. Further, the number of output channels at every encoder/decoder block was reduced to half of the original size used in 3D U-net. For example, the two convolutions in encoder 1 have number of channels $\{ 16, 32 \}$ instead of $\{ 32, 64 \}$. 

\subsubsection{VoxResNet}
\noindent

\begin{figure}[t]
    \centering
    \includegraphics[width=0.98\linewidth]{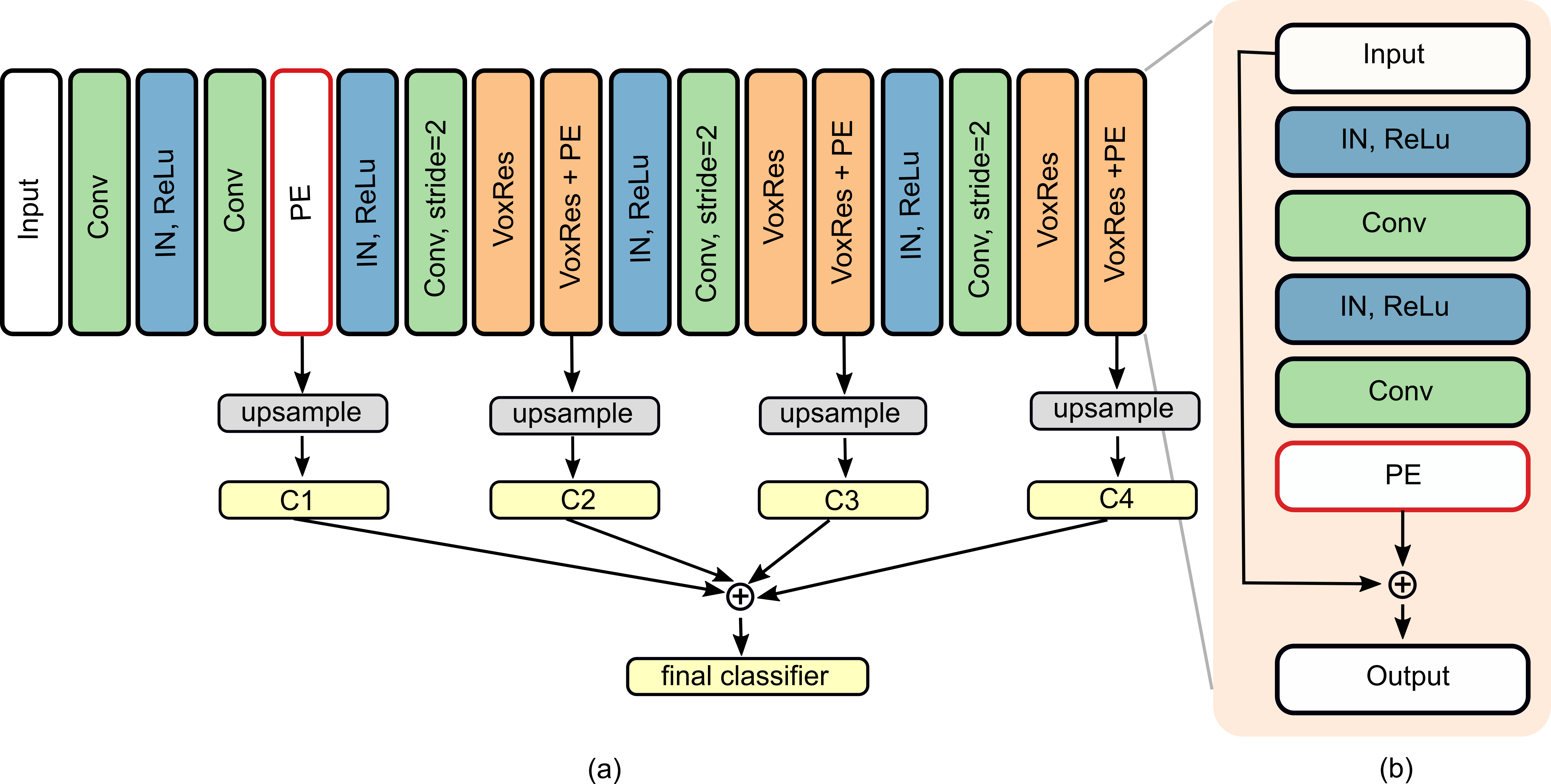}
    \caption{Placement of PE blocks within VoxResNet~\cite{Chen2018} architecture. (a) Illustration of VoxResNet architecture, adapted from~\cite{Chen2018}, with PE blocks added before each downsampling step. (b) PE blocks within VoxRes Module.}
    \label{fig:placement_voxresnet}
\end{figure}

VoxResNet~\cite{Chen2018} is a 3D residual network architecture used for volumetric brain segmentation.
A main building block is the VoxRes module, which is a residual block consisting of two 3D convolutional layers. Downsampling is performed three times, using strided convolutions with a stride of 2. Upsampling is performed using transposed convolutions. The network outputs four auxiliary classifiers and a final classifier which is the sum of all auxiliary classifiers. Since the auxiliary classifiers all have a different receptive field, deep supervision can help with segmenting structures of different sizes. We chose to place the recalibration blocks before each downsampling step. We follow the convention of~\cite{Hu_2018_CVPR} for placement within residual blocks.
The architecture and the placement of recalibration blocks within VoxResNet are illustrated in Fig.~\ref{fig:placement_voxresnet}.

\subsection{Training Parameters and Implementation details}
Due to the large dimensions of the input volumes, we chose a batch size of 1 for training.
\rebuttal{In preliminary experiments, we found that Batch Normalization with applying running mean during testing leads to noisy validation loss and decreased performance on the test set. Therefore we chose Instance Normalization~\cite{instancenorm} instead. We further found that Instance Normalization works better than Group Normalization for our tasks.}
Optimization was done using SGD with momentum of 0.9. \rebuttal{We trained each model for 120 epochs. The learning rate was initially set to 0.1 and was reduced by a factor of 10 when validation loss plateaued for more than 10 epochs.}
Data augmentation using elastic deformations and random rotations was performed on the training set.
We used a combined Cross-Entropy and Dice loss with the Cross-Entropy loss being weighted using median frequency balancing to tackle the high class-imbalance, similar to~\cite{roy2019quicknat}.
For training VoxResNet, we weighted the loss of the auxiliary classifiers by a weighting factor as proposed in~\cite{Chen2018}. Similar to Chen et al., we initially set the weighting factor to $1$ and decreased it by a factor of $2$ every 10 epochs not going below $0.001$. 
All models were trained on Nvidia Quadro P6000 GPU with 24GB of RAM or Nvidia TitanXP GPU with 12 GB of RAM. For training on TitanXP, we used the PyTorch checkpoint functionality, which saves memory by not saving any intermediate activations during the forward pass, but rather recomputing the activations during the backward pass. Using the checkpoint functionality comes with an increase in computation time, but it is useful when training deep networks on volumetric inputs, which requires a large amount of RAM.

\subsection{Evaluation metrics}
To evaluate the segmentation performance, we chose two different evaluation metrics. We use the volumetric Dice similarity coefficient (DSC) as a measure of overlap of the segmentation masks and the surface Dice coefficient as a measure of surface distances.
The surface Dice coefficient~\cite{Ron18a} computes the overlap of the surfaces of two segmentation masks given a specific boundary tolerance. Where the volumetric Dice coefficient is insensitive to small segmentation mistakes in boundary regions, especially for large organs, the surface Dice penalizes these small mistakes. 
The authors in~\cite{Ron18a} propose to set the tolerance parameter specific for each structure depending on previous experiments with several radiologists. Since we did not have several manual segmentations to compute these values, we chose to set the tolerance parameter to the lowest possible value, the resolution of the scans, for all structures.

\section{Results and Discussion}
\label{sec:dis}
The experiments for evaluating the performance of PE blocks are structured as follows. 
First, we verify the architectural choices of PE blocks.
Second, we compare the performance of PE with our previously introduced 3D extensions of existing recalibration methods.
The first two experiments are conducted on the MALC dataset for brain segmentation. 
In the third experiment, we deploy all trained models on two different brain datasets, ADNI and CANDI, to evaluate the performance on unseen data from different age groups and pathologies. 
Fourth, we evaluate whether PE blocks can be used across different segmentation tasks by performing whole-body segmentation on the Visceral dataset without changing any hyperparameters.
In all experiments so far, we use 3D U-net~\cite{cciccek20163d} as our baseline architecture. 
In the final experiment, we integrate PE blocks into VoxResNet and evaluate the performance on MALC dataset.
\begin{table}[t]
    \centering
     \caption{Comparison of pooling methods within the PE block and aggregation strategies to combine the projection vectors. Reported are the mean and standard deviation of volumetric Dice coefficients. The compared modules were integrated into 3D U-net, trained and tested on MALC dataset.}
     \ra{1.2}
    \begin{tabular}{@{}l c c c@{} }
        \toprule
         &  \multicolumn{3}{c}{Pooling}\\
        \cmidrule(lr{.75em}){2-4}
        Aggregation & Avg & Max & Avg\&Max\\
        \midrule
         Add & $\textbf{0.854 $\pm$ 0.075}$ & $0.819 \pm 0.194$ & $0.848 \pm 0.075$  \\
         Max & $0.853 \pm 0.075$ & $0.820 \pm 0.175$ & $0.817 \pm 0.088$\\
         Mult & $0.844 \pm 0.101$ & $0.798 \pm 0.176$ & $0.808 \pm 0.164$\\
         \bottomrule
    \end{tabular}
    \label{tab:pooling}
\end{table}{}

\begin{table*}[!t]
\centering
\ra{1.2}
\caption{Comparison of segmentation performance of 3D U-net on MALC test set with different squeeze and excite based attention blocks and our proposed PE block. Volumetric and surface Dice scores, averaged over the hemispheres, for selected classes. WM = white matter, GM = grey matter, Inf.LV = inferior lateral ventricle, Amygd. = Amygdala and Acc. = Accumbens}
\resizebox{\linewidth}{!} {
    \begin{tabular}{@{}l c c c c c c c c c c c c c@{}}
        \toprule
         &  \multicolumn{6}{c}{\bfseries Volumetric Dice} & \phantom{a}& \multicolumn{6}{c}{\bfseries Surface Dice}\\
         \cmidrule{2-7} \cmidrule{9-14}
         & Mean $\pm$ std &  WM & GM & Inf.LV & Amygd. & Acc. && Mean $\pm$ std &  WM & GM & Inf.LV & Amygd. & Acc.\\
         \midrule
         3D U-net \cite{cciccek20163d} & $0.823 \pm \rebuttal{0.142}$ & $0.918$ & $0.904$&  $0.382$ & $0.785$ & $0.529$ && $0.928 \pm \rebuttal{0.078}$ & $0.981$ & $0.975$ & $0.632$ & $0.921$ & $0.877$ \\
         \hdashline
         3D cSE \cite{Hu_2018_CVPR,zhu2019anatomynet} & $0.845 \pm \rebuttal{0.102}$ & $0.920$ & $\textbf{0.907}$ & $0.488$ &  $0.787$ & $0.754$ && $0.938 \pm \rebuttal{0.061}$ & $0.981$ & $0.975$ & $0.704$ & $0.920$ & $0.943$\\
         3D sSE \cite{roy2019recalibrating} & $0.849 \pm \rebuttal{0.077}$ & $0.918$& $0.904$& $0.618$& $\textbf{0.795}$&$0.751$ && $0.946 \pm \rebuttal{0.022}$ & $0.979$ & $0.973$ & $0.890$ & $0.927$ & $0.939$\\
         3D scSE \cite{roy2019recalibrating} & $0.835 \pm \rebuttal{0.115}$ &  $0.919$& $0.905$&$0.554$&$0.794$&$0.527$ && $0.933 \pm \rebuttal{0.076}$ & $\textbf{0.982}$ & $\textbf{0.976}$ & $0.805$ & $\textbf{0.938}$ & $0.669$ \\
         3D CBAM \cite{woo2018cbam} & $0.831 \pm \rebuttal{0.125}$ & $0.918$& $0.903$&$0.488$&$0.792$&$0.525$ &&  $0.921 \pm \rebuttal{0.088}$ & $0.978$ & $0.971$ & $0.709$ & $0.925$ & $0.661$\\
         Project \& Excite & $ \textbf{0.854 $\pm$ \rebuttal{0.075}}$ &  $\textbf{0.921}$ & $ 0.906$ & $ \textbf{0.627}$ & $0.794$ & $\textbf{0.757}$ && $\textbf{0.951 $\pm$ \rebuttal{0.022}}$ & $0.981$ & $0.975$ & $\textbf{0.893}$ & $0.929$ & $\textbf{0.948}$\\
         \bottomrule
    \end{tabular}}
\label{tab:main_results}
\end{table*}
\subsection{Architecture and Hyperparameters}
\subsubsection{Pooling and Aggregation strategy of PE blocks}
We performed experiments to investigate the choice of pooling strategy and aggregation of the projection vectors. For the projection operation, we compared average pooling with max-pooling and a combination of both pooling strategies. For the combined pooling method, we perform average pooling as described in Sec.~\ref{sec:methods:PE} and separately perform three max-pooling operations along the different dimensions. The obtained average and max projection vectors are then broadcasted to original feature map size and separately passed through the shared convolutional layers. We combine the two recalibration maps by element-wise summation, before passing the recalibration map through the sigmoid layer. 
Furthermore, we evaluated different aggregation strategies to combine the three different projection vectors. Here we compared adding with element-wise max operation and element-wise multiplication.
Tab.~\ref{tab:pooling} shows the results for these experiments. We observe, that average pooling with addition or element-wise max operation as aggregation strategy leads to the best performances. 
We choose addition as the aggregation strategy since it can be computed in place and has lower computational complexity.

\subsubsection{Hyperparameter r}
\label{section:reduction_factor}
The hyperparameter \textit{r} controls the reduction of the channel dimension within the Excitation operator, as described in~\ref{sec:methods:PE}. We compared the performance of 3D U-net with integrated PE blocks on MALC dataset for different values of \textit{r}. We set $r$ to values $\{2,4,8,16\}$ and found that $r=8$ leads to best results. 
\rebuttal{We observed similar behaviour for 3D cSE, sSE and CBAM blocks and therefore set $r=8$ for these blocks as well. For 3D scSE, we set $r=2$, since it lead to better performance.}

\subsubsection{Position of Project \& Excite blocks}
\label{section:position}
In this section, we investigate the optimal position at which the Project \& Excite (PE) blocks should be placed within the F-CNN architecture. We use 3D U-net in our experiment here. We explore $6$ different configurations for the placement of PE blocks. They are i) after every encoder block (P1), ii) after every decoder block (P2), iii) after the bottleneck block (P3), iv) after all encoder and decoder blocks (P4), v) after each encoder block and bottleneck (P5), and finally vi) after all the encoder/ decoder and bottleneck blocks (P6). 
We present the results of all six configurations in Tab.~\ref{tab:placement_of_se} and compare against the baseline 3D U-net model. 
\rebuttal{Firstly, we observe placing PE blocks after encoder (P1) and bottleneck (P3) blocks provides an increase of 1 percentage point whereas placing them after decoder (P2) does not affect the performance.
Secondly, we observe that placing PE blocks after every encoder and decoder block (P4) increases the DSC by $0.026$.} This indicates the fact that PE blocks at decoder have a positive effect when encoder blocks also have PE blocks (contrasting P1, P2, and P4). \rebuttal{Also, we observe that placing PE blocks after encoders and bottleneck (P5) provides a boost in DSC by $0.018$.} This indicates that PE blocks in encoder and bottleneck work better in conjunction (contrasting P1, P3, and P5).
\rebuttal{Finally, by placing PE blocks after all blocks (P6) we observe a boost of $0.03$ in DSC which is higher than the rest of the configurations.} Thus we use this configuration for our experiments. \rebuttal{We further investigated if placing the PE blocks after each convolutional layer within the encoder, decoder, and bottleneck, but did not observe an increase in performance.}

\begin{table}[h]
    \centering
    \ra{1.2}
    \caption{Mean Dice score on MALC dataset due to placement of PE blocks within 3D U-net.}
    \begin{tabular}{@{}l c c c c@{}}
        \toprule
        & \multicolumn{3}{c}{Position of PE block}&\\
        \cmidrule{2-4}
         & Encoders & Bottleneck & Decoders & Mean Dice $\pm$ std \\
        \midrule
         3D U-net & \xmark  & \xmark & \xmark & $0.823 \pm 0.142$\\ \hdashline
         P1 & \cmark & \xmark & \xmark & $0.837 \pm 0.127$\\
         P2 & \xmark & \xmark & \cmark & $0.825 \pm 0.148$\\
         P3 & \xmark & \cmark & \xmark & $0.835 \pm 0.115$\\
         P4 & \cmark & \xmark & \cmark & $0.849 \pm 0.088$\\
         P5 & \cmark & \cmark & \xmark & $0.841 \pm 0.113$\\
         P6 & \cmark & \cmark & \cmark & $ \textbf{0.854 $\pm$ 0.075}$\\
        \bottomrule
    \end{tabular}
    \label{tab:placement_of_se}
\end{table}

\subsection{Comparison of 3D recalibration blocks}

\subsubsection{Structure-wise comparison}
We present the results of whole-brain segmentation in Tab.~\ref{tab:main_results}.
We compare PE blocks to 3D cSE, 3D sSE, 3D scSE, 3D CBAM and the baseline 3D U-net. We were unable to compare to a 3D version of SegSE~\cite{pereira2019adaptive} since computing 3D dilated convolutions on whole volume inputs was not feasible on our GPUs due to the highly increased memory requirement of dilated convolutions.
The placement of the other blocks in the architecture was kept identical to ours.
We report the volumetric and surface Dice coefficients, where we present the mean Dice score over all classes and Dice scores of some selected classes. Note that for simplicity we averaged the Dice coefficients over both hemispheres. 
\rebuttal{We observe the overall mean Dice score by using 3D cSE and 3D sSE increases by $0.02$, whereas PE blocks lead to an increase of $0.03$, substantiating its efficacy. 
Interestingly, the modules that combine channel and spatial recalibration (CBAM and scSE) only lead to an improvement of $0.01$, indicating their 3D version might not be as efficient as the corresponding 2D versions.}
Further, we explored the impact of PE blocks on some selected structures. 
Firstly, we selected bigger structures, white and grey matter.
The boost in Dice score for white and grey matter was marginal for all blocks. 
Next, we analyze some smaller structures, namely inferior lateral ventricles, amygdala, and accumbens, which are difficult to segment.
\rebuttal{We observe an immense boost in Dice score using PE blocks and sSE blocks in these structures ranging from $0.03-0.24$.} cSE, scSE and CBAM blocks also boost the performance but do not reach the performance of PE and sSE blocks. 
\rebuttal{In conclusion, we observe the best performance for PE and 3D sSE models, where the increase in performance for large structures is modest, but for smaller classes, adding these modules can lead to an immense performance boost.}

\begin{table}[ht]
    \centering
    \ra{1.2}
    \caption{Comparison of 3D U-net with integrated recalibration blocks vs. the addition of more convolutional layers. Reported are mean volumetric Dice score and model complexity measured in the increase of number of model parameters, \rebuttal{maximum GPU RAM occupation during training and average inference time (forward pass and calculation of evaluation metrics). Time and memory complexity were measured on a single Titan XP GPU.}}
    \begin{tabular}{@{}l c c c c  @{}}
        \toprule
        & Mean Dice $\pm$ std & \# Params & \rebuttal{Memory} & \rebuttal{Time}\\
        \midrule
        3D-Unet \cite{cciccek20163d} & $0.823 \pm 0.142$ & $5.57 \cdot 10^6$ & \rebuttal{6.7 GB} & \rebuttal{$0.56 s$}\\
        \hdashline
        + 3D cSE \cite{Hu_2018_CVPR,zhu2019anatomynet} & $0.845 \pm 0.102$ &  $ + 0.50 \%$& \rebuttal{7.6 GB} & \rebuttal{$0.85 s$}\\
        + 3D sSE \cite{roy2019recalibrating} & $0.849 \pm 0.077$ & $+ 0.01 \%$& \rebuttal{7.7 GB} & \rebuttal{$0.56 s$}\\
        + 3D scSE \cite{roy2019recalibrating} & $0.835 \pm 0.115$ & $+1.98 \%$& \rebuttal{8.7 GB} & \rebuttal{$0.87 s$}\\
        + 3D CBAM \cite{woo2018cbam} & $0.831 \pm 0.125$ & $+ 0.50 \%$& \rebuttal{8.2 GB} & \rebuttal{$1.19 s$}\\
        + Project \& Excite & $\textbf{0.854 $\pm$ 0.075}$ & $ + 0.50 \%$& \rebuttal{8.7 GB} & \rebuttal{$0.79 s$}\\
        \hdashline
        + Encoder/Decoder & $0.849 \pm 0.086$ & $+ 39.7 \%$ & \rebuttal{6.8 GB} & \rebuttal{$0.60 s$}\\
        + 2 Conv layers & $0.839 \pm 0.115$ & $+ 3.97 \%$ & \rebuttal{6.7 GB} & \rebuttal{$0.57 s$}\\
        \bottomrule
    \end{tabular}
    \label{tab:complexity}
\end{table}

\subsubsection{Model Complexity}
Here we investigate the increase in model complexity due to the addition of PE blocks within 3D U-net architecture. We compare the PE blocks with 3D cSE~\cite{zhu2019anatomynet}, 3D sSE, 3D scSE, and 3D CBAM in Tab.~\ref{tab:complexity}. We present results on the MALC dataset. We observe that even though PE blocks, CBAM and 3D cSE blocks cause the same fraction of $0.5\%$ increase in model complexity, PE blocks provide a higher increase of accuracy at the same expense.
3D sSE has the smallest increase in complexity but does not reach the performance of PE blocks.
Note that 3D scSE blocks lead to a higher increase in parameters due to a lower reduction factor of $2$, as described in Sec.~\ref{section:reduction_factor}.
One might argue that the boost in performance is due to the added complexity, which might also be gained by adding more convolutional layers. 
We investigated this matter by conducting two more experiments. First, we added an extra encoder and decoder block within the architecture. This immensely increased the model complexity by almost $40\%$ and leads to the same performance as adding 3D sSE blocks. Next, we only added two additional convolutional layers at the second encoder and second decoder to make sure that the increase in model complexity is only marginal ($\sim4\%$). Here, we observed an increase in performance similar to scSE with double the increase in parameters but still failed to match the performance of PE blocks. Thus, we conclude that recalibration blocks are more effective than simply adding convolutional layers.
\rebuttal{We further compare the models with respect to the maximum GPU RAM occupation during training and the time for segmenting one scan. We observe that PE and scSE blocks require more GPU RAM than other modules. Inference time for all modules except CBAM is under $1s$ on a Titan XP GPU. We observed that adding PE modules leads to a faster convergence of models, which could be relevant when GPU time is limited. When stopping the training at 80 epochs, we observe a mean overall DSC of $0.845$ for PE models, compared to $0.796$ for the baseline 3D U-Net. }

\begin{figure}[h!]
    \centering
     \includegraphics[width= \linewidth]{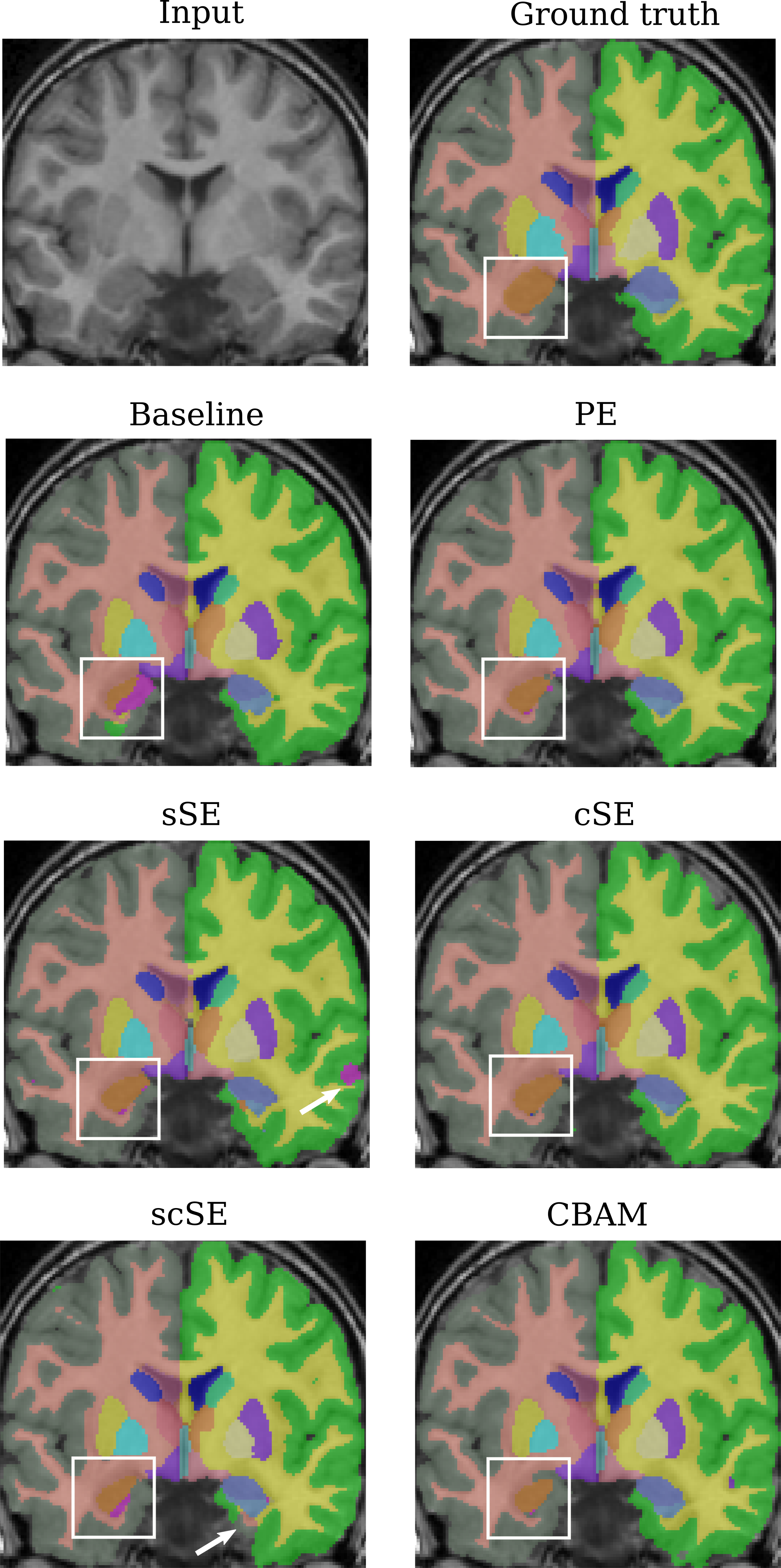}
    \caption{Input scans, manual segmentation, and results for 3D U-net, 3D sSE, cSE, CBAM and our PE model, for CANDI dataset. 
    White boxes indicate the regions where the recalibration blocks improved the performance over the baseline model significantly. The white arrows point to false segmentations in the sSE and scSE models.}
    \label{fig:segmentations_brain}
\end{figure}

\begin{table}[h]
    \centering
    \caption{Segmentation performances of models trained on MALC dataset measured in mean volumetric and surface Dice scores, tested on ADNI and CANDI datasets.}
    \ra{1.2}
    \begin{tabular}{@{} l c c@{}}
    \toprule
     &  \multicolumn{2}{c}{\bfseries ADNI}\\
     \cmidrule{2-3}
    & Volumetric Dice & Surface Dice \\
    \midrule
         3D U-net & $0.743 \pm \rebuttal{0.146} $& $0.820 \pm \rebuttal{0.112}$\\
         \hdashline
         + 3D cSE & $0.769 \pm \rebuttal{0.101}$ & $0.855 \pm \rebuttal{0.065}$ \\
         + 3D sSE & $0.764 \pm \rebuttal{0.087}$ & $0.851 \pm \rebuttal{0.045}$ \\
         + 3D scSE & $0.750 \pm \rebuttal{0.131}$ & $0.831 \pm \rebuttal{0.094}$ \\
         + 3D CBAM & $0.744 \pm \rebuttal{0.125}$ & $0.837 \pm \rebuttal{0.084}$\\
         + Project \& Excite & $\textbf{0.776 $\pm$ \rebuttal{0.080}} $& $\textbf{0.867 $\pm$ \rebuttal{0.046}}$ \\
    \midrule
    
    &  \multicolumn{2}{c}{\bfseries CANDI}\\
     \cmidrule{2-3}
    & Volumetric Dice & Surface Dice \\
    \midrule
         3D U-net &$0.675 \pm \rebuttal{0.169} $ & $0.723 \pm \rebuttal{0.128}$\\
         \hdashline
         + 3D cSE & $0.703 \pm \rebuttal{0.139} $ & $0.755 \pm \rebuttal{0.090}$ \\
         + 3D sSE & $0.690 \pm \rebuttal{0.123}$ & $0.739 \pm \rebuttal{0.085}$ \\
         + 3D scSE & $0.676 \pm \rebuttal{0.167}$ & $0.726 \pm \rebuttal{0.118}$ \\
         + 3D CBAM & $0.699 \pm \rebuttal{0.151}$ & $0.747 \pm \rebuttal{0.103}$\\
         + Project \& Excite & $\textbf{0.719 $\pm$ \rebuttal{0.126}}$ & $\textbf{0.780 $\pm$ \rebuttal{0.078}}$ \\
    \bottomrule
    
    \end{tabular}
    \label{tab:transfer}
\end{table}{}

\subsection{Deployment on unseen datasets}
In the previous experiments, we trained and tested on data of the same dataset (MALC). In this experiment, we explore a more realistic scenario where the model was trained on MALC and deployed on unseen datasets (ADNI and CANDI). 
In this section, we investigate how the different re-calibration blocks aid in achieving robust performance on unseen datasets.
Tab.~\ref{tab:transfer} presents the overall mean volumetric and surface Dice scores for both unseen datasets.
\rebuttal{The addition of PE blocks lead to the highest increase of $0.03$ in DSC, in comparison to other blocks on both datasets.}

We observe that the addition of PE blocks provides a higher boost in performance in comparison to adding sSE blocks for both datasets. \rebuttal{On the CANDI dataset, this difference is especially large with almost $0.03$.} This indicates the effectiveness of PE blocks over sSE blocks \rebuttal{and shows their performance is more robust, even on unseen data from different data distributions.}
It must be noted that the average Dice score is lower than its value on MALC test set reported in Tab.~\ref{tab:main_results}. We believe this is due to the difference in data distribution across the datasets.
In Fig.~\ref{fig:segmentations_brain}, we present visualizations of the segmentation performance of PE models in comparison to baseline 3D U-net and other recalibration blocks for CANDI dataset.

\subsection{Experiments on whole-body segmentation}
\begin{figure}[h!]
    \centering
     \includegraphics[width=0.92 \linewidth]{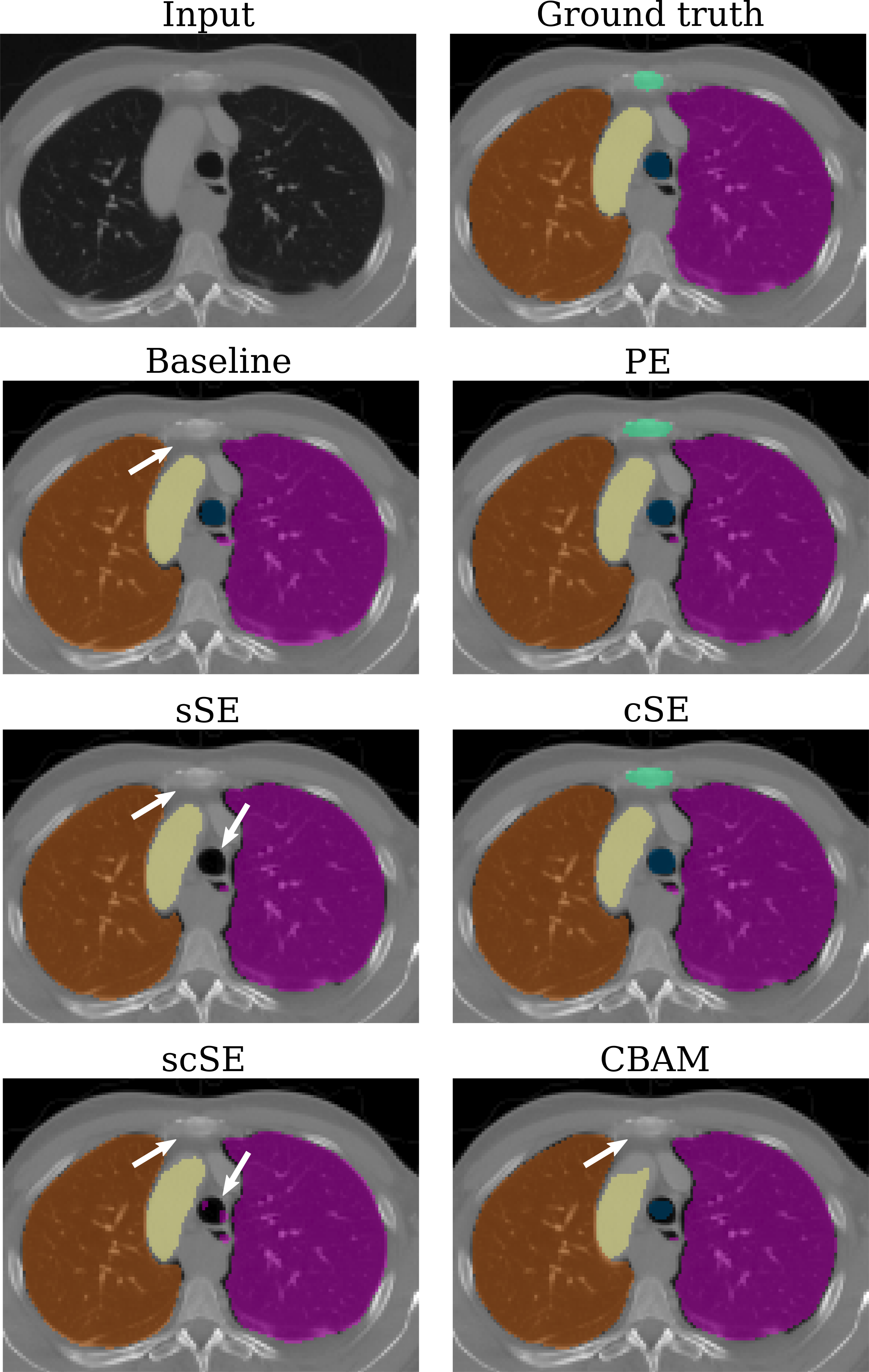}
    \caption{Input scans, manual segmentation, and results for 3D U-net, 3D sSE, cSE, scSE, CBAM and our PE model, for Visceral dataset. 
    The white arrows point to the structures where the models missed segmentations of sternum and trachea.}
    \label{fig:segmentations_visceral}
\end{figure}

To determine if PE blocks generalize to a different task and modality, we evaluate their performance on whole-body segmentation on contrast-enhanced CT scans.
Tab.~\ref{tab:visc_results} reports volumetric and surface Dice scores on the Visceral dataset for all models.
\rebuttal{Contrary to results observed on brain datasets, 3D sSE performs worst on visceral dataset decreasing the baseline mean Dice score by almost $0.06$.} 
scSE and CBAM also do not reach the performance of the baseline model. 
When looking at selected bigger structures,  liver and right lung, we observe a similar trend to brain segmentation, where the performance of all models is comparable to the baseline 3D U-net.
Next, we analyze some smaller structures, namely the right kidney, trachea, and sternum, which are more difficult to segment.
The increase of the Dice score in kidneys and trachea are rather small for both cSE and PE models. CBAM leads to the highest performance for trachea, but the overall performance of this model is poor.
\rebuttal{We see a high improvement of $\sim0.3$ in DSC for sternum in both cSE and PE models, whereas the sSE model fails to segment the sternum completely. The performance of 3D scSE and CBAM is also poor on this class.} 
Since all of these modules include squeezing of the channel dimension, this indicates the importance of information encoded in the channel dimension.
We conclude that PE models lead to the best overall results and also give the most consistent performance over all structures, \rebuttal{which indicates that PE blocks are more robust.}
We present visualizations of all segmentations in Fig.~\ref{fig:segmentations_visceral}. We show a slice of the thorax with segmentation of lungs, aorta, trachea, and sternum, where the baseline model, sSE, scSE, and CBAM model fail to segment the sternum, and sSE and scSE models also fail to segment the trachea.

\begin{table*}[ht!]
\centering
\ra{1.2}
\caption{Comparison of segmentation performance of 3D U-net on Visceral dataset with recalibration blocks. Volumetric and surface Dice scores for selected classes. For lungs and kidneys, the right side is reported.}
\resizebox{\linewidth}{!} {
    \begin{tabular}{@{}l c c c c c c c c c c c c c@{}}
        \toprule
         &  \multicolumn{6}{c}{\bfseries Volumetric Dice} & \phantom{a}& \multicolumn{6}{c}{\bfseries Surface Dice}\\
         \cmidrule{2-7} \cmidrule{9-14}
         &Mean &Liver & Lung & Kidney & Trachea & Sternum && Mean &Liver & Lung & Kidney & Trachea & Sternum  \\
         \midrule
         3D U-net \cite{cciccek20163d} & $0.810 \pm \rebuttal{0.137}$ &$0.922$ & $0.965$ & $0.907$ & $0.815$ & $0.438$ && $0.771 \pm \rebuttal{0.121}$&$0.755$&$0.924$&$0.857$&$0.895$&$0.481$ \\
         \hdashline
         + 3D cSE \cite{Hu_2018_CVPR,zhu2019anatomynet} &  $0.837 \pm \rebuttal{0.091}$  & $0.929$ & $\textbf{0.967}$ & $0.916$ & $0.817$ & $\textbf{0.737}$ && $0.805 \pm \rebuttal{0.092}$&$0.776$&$\textbf{0.936}$&$0.880$&$0.900$&$\textbf{0.799}$\\
         + 3D sSE \cite{roy2019recalibrating}& $0.751 \pm \rebuttal{0.268}$ & $0.925$ &$0.964$& $0.919$&$0.347$&$0$&& $0.711 \pm \rebuttal{0.248}$&$0.768$&$0.915$&$\textbf{0.896}$&$0.381$&$0$\\
         + 3D scSE \cite{roy2019recalibrating} & $0.802 \pm \rebuttal{0.147}$ & $0.927$ &$0.966$& $0.914$&$0.659$&$0.419$&& $0.763 \pm \rebuttal{0.125}$&$0.766$&$0.933$&$0.882$&$0.729$&$0.454$\\
         + 3D CBAM \cite{woo2018cbam} & $0.797 \pm \rebuttal{0.174}$ & $0.924$ &$0.954$& $0.913$&$\textbf{0.831}$&$0.291$&& $0.752 \pm \rebuttal{0.156}$&$0.750$&$0.901$&$0.875$&$0.902$&$0.316$\\
         + Project \& Excite & $\textbf{0.844 $\pm$ \rebuttal{0.088}}$  &$\textbf{0.934}$ & $\textbf{0.967}$ & $\textbf{0.920}$ & $0.822$ & $0.733$ && $\textbf{0.814 $\pm$ \rebuttal{0.086}}$&$\textbf{0.779}$&$0.934$&$0.895$&$\textbf{0.905}$&$0.796$\\
         \bottomrule
    \end{tabular}}
\label{tab:visc_results}
\end{table*}

\begin{table*}[ht]
\centering
\ra{1.2}
\caption{Comparison of segmentation performance of VoxResNet on MALC test set with different recalibration blocks. Volumetric and surface Dice scores, averaged over the hemispheres, for selected classes. WM = white matter, GM = grey matter, Inf.LV = inferior lateral ventricle, Amygd. = Amygdala and Acc. = Accumbens}
\resizebox{\linewidth}{!} {
    \begin{tabular}{@{}l c c c c c c c c c c c c c@{}}
        \toprule
         &  \multicolumn{6}{c}{\bfseries Volumetric Dice} & \phantom{}& \multicolumn{6}{c}{\bfseries Surface Dice}\\
         \cmidrule{2-7} \cmidrule{9-14}
         & Mean &  WM & GM & Inf.LV & Amygd. & Acc. && Mean &  WM & GM & Inf.LV & Amygd. & Acc.\\
         \midrule
         VoxResNet \cite{Chen2018} & $0.855 \pm \rebuttal{0.076}$ & $0.922$ & $0.908$&  $0.621$ & $0.779$ & $0.769$ && $0.938 \pm \rebuttal{ 0.022}$ & $0.933$ & $0.939$ & $0.895$ & $0.909$ & $0.951$ \\
         \hdashline
         + 3D cSE \cite{Hu_2018_CVPR,zhu2019anatomynet} & $0.859 \pm \rebuttal{0.071}$ & $0.926$ & $0.912$ & $0.653$ &  $0.779$ & $0.768$ && $0.942 \pm \rebuttal{0.021}$ & $0.941$ & $0.946$ & $0.903$ & $0.911$ & $\mathbf{0.952}$\\
         + 3D sSE \cite{roy2019recalibrating} & $0.852 \pm \rebuttal{0.072}$ & $0.916$& $0.903$& $0.644$& $0.775$&$0.758$ && $0.934 \pm \rebuttal{0.022}$ & $0.920$ & $0.930$ & $0.898$ & $0.908$ & $0.944$\\
         + 3D scSE \cite{roy2019recalibrating} & $0.828 \pm \rebuttal{0.106}$ &  $0.911$& $0.899$&$0.552$&$0.763$&$0.599$ && $0.908 \pm \rebuttal{0.054}$ & $0.908$ & $0.923$ & $0.799$ & $0.895$ & $0.752$ \\
         + 3D CBAM \cite{woo2018cbam} & $0.853 \pm \rebuttal{0.070}$ & $0.919$& $0.905$&$0.652$&$0.781$&$0.759$ &&  $0.934 \pm \rebuttal{0.022}$ & $0.926$ & $0.933$ & $0.901$ & $0.916$ & $0.942$\\
         + Project \& Excite & $ \textbf{0.861 $\pm$ \rebuttal{0.072}}$ &  $\textbf{0.941}$ & $ \textbf{0.926}$ & $ \textbf{0.657}$ & $\textbf{0.789}$ & $\textbf{0.771}$ &&$\textbf{0.947 $\pm$ \rebuttal{0.023}}$ & $\textbf{0.984}$ & $\textbf{0.977}$ & $\textbf{0.904}$ & $\textbf{0.922}$ & $\textbf{0.952}$\\
         \bottomrule
    \end{tabular}}
\label{tab:voxresnet}
\end{table*}

\subsection{Experiments on VoxResNet}
\label{section:exp_voxresnet}
After investigating the performance of PE blocks for two different tasks and across multiple datasets, we evaluated the effectiveness of PE blocks when integrated into a different architecture. We select VoxResNet~\cite{Chen2018} for this purpose, with results on MALC dataset presented in Tab.~\ref{tab:voxresnet}.
\rebuttal{We observe that VoxResNet outperforms 3D U-net by $0.03$ in DSC on average.} This is mainly due to a higher performance for small structures like inferior lateral ventricle and accumbens. We believe this is achieved due to the deep supervision which makes the VoxResNet architecture better suited for segmentation of small structures compared to 3D U-net.
We observe an increase in volumetric and surface Dice scores using PE blocks for all structures. 3D csE also increases the performance, but the other blocks 3D sSE, 3D scSE, and 3D CBAM lead to an overall decrease in performance. 
\rebuttal{When looking at larger structures, white matter and grey matter, the Dice score increases by $0.02$ when using PE blocks, in contrast to 3D U-net, where the performance for large structures was similar to the baseline model.} The performance of the other recalibration blocks is very close to the baseline model for larger structures.
For small structures, PE blocks lead to an increase in performance, although it is not as significant as in 3D U-net. Interesting to see is that 3D scSE blocks lead to a decrease in performance on all smaller structures, although its components (3D cSE and 3D sSE) perform well. \rebuttal{This could be due to the different reduction factor $r$ for scSE.}

\comment{
\begin{figure*}[h]
    \centering
     \includegraphics[width=\linewidth]{figures/segm.png}
    \caption{Input scans, manual segmentation, and results for 3D U-net, 3D sSE and our PE model, for CANDI (top row) and Visceral (bottom row) datasets. 
    In the top row, white boxes indicate the regions where PE and sSE models improved the performance over the baseline model significantly. The white arrow points to a false segmentation in the sSE model.
    In the bottom row, white arrows point to the structures where the models missed segmentations of sternum and trachea.}
    \label{fig:segmentations}
\end{figure*}}

\section{Conclusion}
\label{sec:conc}
In this work, we focused on the task of whole volume medical image segmentation using 3D F-CNNs and targeted the challenges specific to it.
Due to the added dimensionality 3D F-CNNs are often used with limited depth and limited features to keep model complexity under control. We explore the usage of feature recalibration to boost their performance. 
First, we provided 3D extensions of multiple existing 2D recalibration techniques.
Following, we presented the generic `compress, process, recalibrate' framework for easy comparison of all recalibration blocks. 
Finally, we proposed Project \& Excite (PE), a light-weight recalibration module custom made for 3D F-CNN architectures, which boosts segmentation performance while increasing model complexity by a small fraction. 
In exhaustive experiments on multiple datasets and multiple applications, we demonstrated that PE blocks do not only provide better recalibration in comparison to other blocks but are also more efficient than simply adding more convolutional layers in 3D F-CNNs.
One interesting finding is that PE modules lead to a higher boost in segmentation performance for small structures in contrast to other recalibration blocks. 
\rebuttal{We believe, this is due to the retained spatial information within the project operation.}
\rebuttal{We observed that PE blocks consistently perform well, on different datasets and different base architectures, whereas other recalibration blocks sometimes even decrease the performance.
We conclude PE blocks are a good and robust design choice for 3D segmentation tasks, especially when the target structures are small.
}

\section*{Acknowledgments}
This research was partially supported by the Bavarian State Ministry of
Science and the Arts in the framework of the Centre Digitisation.Bavaria (ZD.B). We thank NVIDIA Corporation for GPU donation.

\bibliographystyle{IEEEtran}

\bibliography{references}

\end{document}